\documentstyle[preprint,aps]{revtex}
\begin{document}
\draft

\title{Tunneling and Quantum Noise in 1-D Luttinger Liquids}
\author{C. de C. Chamon$^1$, D. E. Freed$^2$ and X. G. Wen$^1$}
\address{$^1$Department of Physics, Massachusetts
Institute of Technology, \\ Cambridge, Massachusetts 02139}
\address{$^2$Center for Theoretical Physics \\ Laboratory for Nuclear
Science \\ and Department of Physics, Massachusetts
Institute of Technology, \\ Cambridge, Massachusetts 02139}
\maketitle
\begin{abstract}
We study non-equilibrium noise in the transmission current through
barriers in 1-D Luttinger liquids and in the tunneling current between
edges of fractional quantum Hall liquids. The distribution of
tunneling events through narrow barriers can be described by a Coulomb
gas lying in the time axis along a Keldysh (or non-equilibrium)
contour.  We show that the charges tend to reorganize as a dipole gas,
which we use to describe the tunneling statistics. The dipole-gas
picture allows us to have a unified description of the low frequency
shot noise and the high frequency ``Josephson'' noise. The correlation
between the charges within a dipole (intra-dipole) contributes to the
high-frequency ``Josephson'' noise, which has an algebraic singularity
at $\omega=e^*V/\hbar$, whereas the correlations between dipoles
(inter-dipole) are responsible for the low-frequency noise. We show
that an independent or non-interacting dipole approximation gives a
Poisson distribution for the locations of the dipole centers of mass,
which gives a flat noise spectrum at low-frequencies and corresponds
to uncorrelated shot noise.  Including inter-dipole interactions gives
an additional $1/t^2$ correlation between the tunneling events that
results in a $|\omega|$ singularity in the noise spectrum. We present
a diagrammatic technique to calculate the correlations in perturbation
theory, and show that contributions from terms of order higher than
the dipole-dipole interaction should only affect the strength of the
$|\omega|$ singularity, but its form should remain $\sim |\omega|$ to
all orders in perturbation theory. A counting argument also suggests
that the leading algebraic singularity at $\omega_J$ should be
$\propto |\omega-\omega_J|^{2g-1}$ to all orders in perturbation
theory.
\end{abstract}


\section{Introduction}

The noise spectrum in a two-terminal conductor in the absence of an
applied voltage is proportional to the conductance and to the
temperature. This result was found experimentally by Johnson in
1927 \cite{Johnson} and explained theoretically by Nyquist in 1928
\cite{Nyquist}. Such relation between equilibrium noise and
conductance can be seen as a consequence of the
fluctuation-dissipation theorem. The noise in the presence of
transport (non-equilibrium noise) can also be related to transport
coefficients for non-interacting systems \cite{Lesovik,Buttiker}, but now
these transport coefficients, in the most general case, cannot be
determined from conductance measurements alone.  For interacting
systems one should expect an even richer behavior, as new features in
the noise should appear as a consequence of correlations due to
interactions. In general the shape of the noise spectrum is determined
by the dynamical properties of the system, which contain information
about the excited states. Thus the noise spectrum is a powerful probe
which allows us to study dynamics of strongly correlated systems.

Interacting electron systems at 1-D form strongly correlated states
--Luttinger liquids-- whose properties are well understood.  However,
it has been very difficult to realize 1-D Luttinger liquids in
experiments.  This is because even a small amount of impurities cause
the localization of the 1-D electrons and destroy the Luttinger
liquids. Recently it was realized that another strongly correlated 1-D
state -- Chiral Luttinger liquid -- exists on the edges of fractional
quantum Hall (FQH) liquids. Due to its chirality ({\it i.e.} all
excitations move in the same direction) and the lack of back
scattering, a chiral Luttinger liquid cannot be localized by
impurities. Thus it is possible to realize, in practice, extended 1-D
systems through FQH states.  Recently an IBM group \cite{Webb} studied
the tunneling between two edges of filling fraction $1/3$ FQH states.
They found that the tunneling conductance has a power law dependence
on temperature which is a characteristic property of (chiral)
Luttinger liquids \cite{XGW2,Kane&Fisher1,F&N,CCC&XGW}.  Their finding
is consistent with the theoretical prediction $\sigma
\propto T^4$ for the $\nu=1/3$ FQH state \cite{XGW2}. In this paper we will
study the noise spectrum in the tunneling current between (chiral)
Luttinger liquids. The noise spectrum carries rich information about
dynamical properties of (chiral) Luttinger liquids, which will help us
identify such strongly correlated states in experiments.

Recent studies of noise in non-interacting systems reveal that the
noise spectrum contain features that come from the statistics of the
tunneling particles \cite{Levitov&Lesovik}.  These
statistics-dependent features are not contained in the DC conductance.
For Luttinger liquids, the tunneling particles sometimes carry
fractional statistics and fractional charges. It is then very
interesting to study the noise spectrum for tunneling between (chiral)
Luttinger liquids, especially those features that come from the
strongly correlated properties of (chiral) Luttinger liquids (such as
fractional statistics and fractional charges).

Two kinds of noise may appear in tunneling at a finite voltage $V$,
the shot noise and the ``Josephson'' noise. The shot noise can be
understood from a classical picture in which the average tunneling
current is viewed as a result of many tunneling events. A tunneling
event represents a single particle (which can be an electron or a
charged quasiparticle) that tunnels through the junction. The spectrum
of the shot noise is determined by the correlations between tunneling
events. In this paper we always assume that the tunneling time is much
shorter than the average spacing between two tunneling events. Under
this approximation, we will ignore the retardation and model the
tunneling by an instantaneous tunneling operator $\Gamma
\psi^\dagger_L \psi_R + H.c.$, which transfers particles between two
reservoirs. The ``Josephson'' noise is related to the fact that the
two systems connected by the junction have different chemical
potentials.  The quantum interference between wave functions on the
two sides of the junction may cause a singularity at frequency
$\omega=e^*V/\hbar$ in noise spectrum (here $e^*$ is the charge of the
tunneling particle).  Such features near the ``Josephson'' frequency
$\omega_J\equiv e^*V/\hbar$ are called ``Josephson'' noise.  In this
paper we will develop a language for non-equilibrium noise in 1-D
Luttinger liquids which covers both the shot noise and the
``Josephson'' noise.

We start with the Keldysh formalism, in which the tunneling events are
described by a Coulomb gas of charges on a Keldysh contour. Under
certain conditions the charges at different branches of the contour
pair into dipoles (in this case the Coulomb gas is said to be in the
dipole phase). The dipoles correspond to the tunneling events in the
shot-noise picture. The non-interacting dipole approximation leads to
a Poisson distribution for the separation of dipoles, which results in
a white noise ({\it i.e.} a frequency independent noise) at low
frequencies. However, for a finite voltage across the junction, we find
that the dipoles have a non-zero dipole moment which leads to a long
range $1/t^2$ interaction between dipoles. The dipole-dipole
interaction gives rise to a non-trivial distribution of the tunneling
events which induces a $|\omega|$ singularity in the low frequency
noise spectrum. The dipoles have finite size and the intra-dipole
structures are found to be responsible for the high frequency
``Josephson'' noise, which appear as an algebraic singularity of the
form $|\omega-\omega_J|^{2g-1}$ in the noise spectrum within perturbation
theory.

The full expression for the singularity at zero frequency in the noise
spectrum due to the dipole-dipole interaction is found to be
\begin{equation}
S_{sing.}(\omega)=4\pi g(2g-1)^2(\frac{I_t}{\omega_J})^2\ |\omega|\ ,\nonumber
\end{equation}
where $I_t$ is the average tunneling current and $g$ contains information on
the interactions in the Luttinger liquid (or filling fraction of the
FQH states, in the case of chiral Luttinger liquids).
Because of the nonlinear dependence of $I_t$ on $\omega_J$
\cite{XGW2,Kane&Fisher1,F&N,CCC&XGW}, the strength of the singularity
in the noise spectrum at zero frequency will also have a non-linear
dependence on $\omega_J$
(\ $(\frac{I_t}{\omega_J})^2\propto\omega_J^{4(g-1)}$). The particular
case of non-interacting electrons can be obtained with $g=1$, where
one recovers the $|\omega|$ singularity that appears to order $D^2$ in
the transmission coefficient $D$
\cite{Yang}. The correlations in the case of non-interacting electrons
come from the Pauli principle, which enters very simply in the
formulation used in this paper through the language of bosonization.

The paper is organized as the following.  In section II we will review
the bosonization scheme for 1-D fermionic systems. In section III we
calculate the noise perturbatively. In section IV we use the
non-equilibrium (Keldysh) scattering operator as a means to obtain a
joint probability distribution for tunneling events. The tunneling
events can be mapped into charges of a Coulomb gas, which tend to
reorganize as a dipole gas. A non-interacting dipole approximation
leads to uncorrelated noise. Dipole-dipole interactions and
correlations will be discussed in section V, which lead to an
$|\omega|$ singularity in the low frequency noise spectrum. In section
VI a diagrammatic technique is presented that accounts for the
correlations in a systematic way. We show the existence of the
$|\omega|$ singularity at zero frequency to all orders in perturbation
theory. A counting argument also suggests that the leading singularity
at $\omega_J$ should remain of the form $|\omega-\omega_J|^{2g-1}$ to
all orders in perturbation theory.

\section{Tunneling in 1-D Luttinger Liquids}

In this paper we will study the effect of particle interactions in the
noise spectrum of a 1-D conductor. The results for 1-D systems of
interacting particles, or 1-D Luttinger Liquids, can be directly used
to study noise in the tunneling current between two edge channels in
the Fractional Quantum Hall (FQH) regime. Figure 1 displays the
geometries we are considering here. Fig. 1a shows a 1-D channel
connected to two reservoirs, with a weak link or tunneling barrier in
the middle of the channel. Figs. 1b and 1c show two configurations in
which we can observe tunneling between edge channels. The
configurations can be accessed experimentally using metallic gates
placed on top of the 2-D electron gas. Applying a negative gate
voltage depletes the electron concentration underneath the gate,
causing the two branches of edge states to get closer, and thus
enhancing the tunneling between the channels. Because in this
configuration both edges form the boundary of the same QH liquid,
there can be either electron or quasiparticle (carrying fractional
charge) tunneling. By applying a sufficiently large gate voltage, one
can obtain the situation in Fig. 1c, where the edges form the
boundaries of two disconnected QH liquids, and thus only electrons can
tunnel from one edge to the other.

Both the interacting 1-D systems and the FQH edge states are best
described in the bosonized language \cite{Haldane,XGW1}. In the
case of the interacting 1-D system, the electron operator can be
written as:
\begin{equation}
\psi^{\dagger}\sim\sum_{n\ odd}\ e^{in(k_F x+\theta)}\ e^{i\phi}
\end{equation}
where the $\phi$ and $\theta$ fields satisfy the equal-time
commutation relations
\begin{equation}
[\phi(t,x),\theta(t,y)]=-i\frac{\pi}{2}{\rm sgn}(x-y)\ .
\end{equation}
The canonical momenta associated with $\phi$ and $\theta$ are then
$\Pi_{\phi}=\frac{1}{\pi}\partial_x\theta$ and
$\Pi_{\theta}=\frac{1}{\pi}\partial_x\phi$
respectively. The dynamics of $\phi$ and $\theta$ are described by the
Hamiltonian density
\begin{equation}
{\cal H}=
\frac{1}{2\pi}[g(\partial_x\phi)^2+\frac{1}{g}(\partial_x\theta)^2]\ ,
\end{equation}
where the effect of interactions enters through $g$ \cite{Haldane}.
For repulsive interactions $g<1$, whereas for attractive
interactions $g>1$. For non-interacting electrons $g=1$. The electron
propagator has a power law decay envelope, with the long-range
behavior dominated by $\langle\psi^\dagger(t,0)\psi(0,0)\rangle \propto
t^{-(g+g^{-1})/2}$.

The presence of a weak link or a potential barrier in the channel
gives an additional term in the Hamiltonian which can be expressed in
terms of the bosonic fields $\phi$ and $\theta$ \cite{Kane&Fisher1}.
For a potential barrier located at $x=0$ the perturbation is
$\propto\psi^\dagger(t,x=0)\psi(t,x=0)$, which can be written
(keeping only the most relevant term) as
\begin{equation}
{\cal H}_{int}=\Gamma \delta(x)\ e^{i2\theta(t,0)}+H.c.\ \ .
\end{equation}
For a weak link one can also show that the perturbation is
\begin{equation}
{\cal H}_{int}=\Gamma \delta(x)\ e^{i2\phi(t,0)}+H.c.\ \ .
\end{equation}

Using a rescaling $\tilde{\phi}=2\sqrt{g}\phi$ and
$\tilde{\theta}=\frac{2}{\sqrt{g}}\theta$, the Lagrangian densities
for the small barrier and weak link problems are respectively
\begin{equation}
{\cal L}=\frac{1}{8\pi}[(\partial_t\tilde{\theta})^2-
(\partial_x\tilde{\theta})^2]-
\ \Gamma \delta(x)\ e^{i\sqrt{g}\tilde{\theta}(t,0)}+H.c.
\label{Ltheta}
\end{equation}
and
\begin{equation}
{\cal L}=\frac{1}{8\pi}[(\partial_t\tilde{\phi})^2-
(\partial_x\tilde{\phi})^2]-
\ \Gamma \delta(x)\ e^{i\frac{1}{\sqrt{g}}\tilde{\phi}(t,0)}+H.c.
\label{Lphi}
\end{equation}

Now, for the FQH edge states, we can write the right and left moving
electron and quasiparticle operators as $\Psi_{R,L}(x,t)=:e^{\pm i
\sqrt{g} \phi_{R,L} (x,t)}:\ $, where $g$ is related to the FQH bulk
state. For example, for a Laughlin state with filling fraction
$\nu=1/m$ we have $g=m$ for electrons and $g=1/m$ for
quasiparticles carrying fractional charge $e/m$. The
$\phi_{R,L}$ fields satisfy the equal-time commutation relations
\begin{equation}
[\phi_{R,L}(t,x)\ ,\ \phi_{R,L}(t,y)]=\pm i\pi \ {\rm sgn}(x-y)\ .
\end{equation}
The dynamics of $\phi_{R,L}$ is described by
\begin{equation}
{\cal L}_{R,L}=\frac{1}{4\pi}\ \partial_x\phi_{R,L}
\ (\pm \partial_t- v \partial_x)\phi_{R,L}\ ,
\end{equation}
where $v$ is the velocity of edge excitations (which we will set to 1).
The same algebraic decay of the electron operator occurs in the edge
states of the FQH effect, where we have a chiral Luttinger liquid with
the exponent $g$ directly related to the bulk state (for a review see Ref.
\cite{XGW1}).

The tunneling between left and right moving branches can be written as
$H_{tun}=\Gamma \Psi^\dagger_L \Psi_R + H.c.$. We can write, in terms
of $\phi=\phi_R + \phi_L$, the following total Lagrangian density:
\begin{equation}
{\cal L}=\frac{1}{8\pi}[(\partial_t\phi)^2-
(\partial_x\phi)^2]-
\ \Gamma \delta(x)\ e^{i\sqrt{g}\phi(t,0)}+H.c.\ \ ,\label{L}
\end{equation}
with $\phi$ satisfying
$[\phi(t,x),\partial_t\phi(t,y)]=4\pi i\delta(x-y)$.

The Lagrangian for $\phi$ in Eq. (\ref{L}) is exactly the same as the
one for $\tilde{\theta}$ in Eq. (\ref{Lphi}) and, with $g\rightarrow
1/g$, the same as the one for $\tilde{\phi}$ in Eq. (\ref{Ltheta}).
It is this Lagrangian in Eq. (\ref{L}) that will be the basis of our
work. A voltage difference between the two reservoirs at the ends of
the 1-D channel, or between the edges of the QH liquid, can be easily
introduced in the model by letting $\Gamma \rightarrow \Gamma
e^{-i\omega_0 t}$, where $\omega_0\equiv \omega_J \equiv e^*V/\hbar$,
with $e^*=e$ for electron tunneling and $e^*=e/m$ for quasiparticle
tunneling.

Notice that in order to obtain the coupling term we assume that we
only have contributions from $x=0$ for the tunneling operators. This
is the case when the width of the barrier is narrow. Also, if the
barrier is narrow, the time spent in the tunneling is small
compared to the spacing between tunneling events. Indeed, in this case
we can speak of tunneling events that occur at rather well defined
time coordinates.

Using this language, the average tunneling current through a barrier
in a one dimensional channel and between edge states in the FQH regime
was calculated \cite{XGW2,Kane&Fisher1,F&N,CCC&XGW}. The
current has a nonlinear dependence on the applied voltage between the
terminals, with the power dependence on the voltage intimately related
to the exponent $g$ in the electron propagator. For the case of
tunneling through a single barrier in a 1-D channel, or non-resonant
tunneling between FQH edge states, one finds that $I_t\sim V^{2g-1}$
at zero temperature. In this paper we will study the noise in this
current, starting with a perturbative calculation and then moving to a
formalism that grasps non-perturbative contributions.

\section{Perturbative Calculation}

We can show that the tunneling current operator is
$I_t(t)=j(t)=ie^*\Gamma e^{i\sqrt{g}\phi(t,0)}+H.c.\ $. For example,
in the case of tunneling between edges (such as in Figs. 1b \& 1c) we
simply use that $I_t=-\frac{1}{i\hbar}[N_L,H]=
\frac{1}{i\hbar}[N_R,H]$ (where $N_{R,L}$ are the total charge
operators on the $R,L$ edges) and the commutation relations to obtain
the expression for $I_t$. Similarly, we can find the same for the case
of a 1-D interacting system. The noise spectrum can be obtained by
calculating two-point correlations involving the operator $I_t(t)$.

Notice that, as the problem under consideration is intrinsically
non-equilibrium, one should use the Keldysh (or non-equilibrium)
formalism \cite{Keldysh} in computing expectation values of operators.
This is the case here, where if we treat the coupling term
perturbatively and introduce an adiabatic turning on and off of the
interaction, the state at $t=-\infty$ differs from the one at
$t=\infty$; the charge transfer in one direction due to the applied
voltage clearly makes the two states at $\pm \infty$ different, as the
total charge in one edge branch (or reservoir) decreases whereas in
the other the total charge increases. This problem could, in
principle, be circumvented by including another term in the Hamiltonian
that would close the circuit and bring the charges that tunneled
through the barrier back to the reservoirs (a ``battery''). Such a way
of thought is relevant to clarify the distinction between the
equilibrium and non-equilibrium formalism, and how they can be
connected in principle.  However, in practice, adding the restoring
charge coupling in the Hamiltonian only would make the problem more
cumbersome and poorly defined, which makes the non-equilibrium
formalism a natural choice.

For perturbative calculations of zeroth and first order, however,
there is no difference between the results for expectation values
obtained with either the equilibrium or the non-equilibrium formalism.
This is the case in the calculation of the current-current
correlation, where the lowest order contribution is the zeroth order:
\begin{equation}
\langle j(t)j(0)\rangle ={e^*}^2\langle 0|(i\Gamma e^{-i\omega_0 t}
e^{i\sqrt{g}\phi(t)}-
i\Gamma^* e^{i\omega_0 t} e^{-i\sqrt{g}\phi(t)})(i\Gamma e^{i\sqrt{g}\phi(0)}-
i\Gamma^* e^{-i\sqrt{g}\phi(0)})|0\rangle\ .
\end{equation}
The non-zero contributions come from the terms that, when applied to
$|0\rangle$, transfer zero total charge, so we can write
\begin{eqnarray}
\langle j(t)j(0)\rangle &=&{e^*}^2|\Gamma|^2 \left( e^{-i\omega_0 t}
\langle 0|e^{i\sqrt{g}\phi(t)}
e^{-i\sqrt{g}\phi(0)}|0\rangle + e^{i\omega_0 t} \langle
0|e^{i\omega_0 t} e^{-i\sqrt{g}\phi(t)} e^{i\sqrt{g}\phi(0)}|0\rangle
\right) \nonumber\\ &=&{e^*}^2|\Gamma|^2 \ (e^{-i\omega_0 t} + e^{-i\omega_0
t})
\ e^{g\langle 0|\phi(t)\phi(0)|0\rangle}\ .
\end{eqnarray}
The $\phi$ field correlation is $\langle
0|\phi(t)\phi(0)|0\rangle=-2\ln (\delta +it)$, where $\delta$ is an
ultraviolet cut-off scale. The current-current correlation is then given by
\begin{equation}
\langle j(t)j(0)\rangle ={e^*}^2|\Gamma|^2\ \frac{2\cos(\omega_0 t)}{(\delta
+it)^{2g}}\ ,
\end{equation}
which displays clearly oscillations at frequency $f=\omega_0/2\pi=e^*V/h$.
This implies that the noise spectrum will also display structure at
this frequency. The noise spectrum is calculated from the current-current
correlation:
\begin{eqnarray}
S(\omega)&=&\int_{-\infty}^\infty dt\ e^{i\omega t}
\langle \{ j(t),j(0)\}\rangle \nonumber\\
&=&{e^*}^2|\Gamma|^2
[c_+(\omega_0 +\omega)+c_-(\omega_0 +\omega)
+c_+(\omega_0 -\omega)+c_-(\omega_0 -\omega)]\ ,
\end{eqnarray}
where
\begin{equation}
c_{\pm}(\omega)= \int_{-\infty}^\infty dp \frac{e^{-i\omega p}}{(\delta \mp
ip)^{2g}}
=\frac{2\pi}{\Gamma(2g)}|\omega|^{2g-1}e^{-|\omega|\delta} \ \theta (\pm\omega)
\ .
\end{equation}
The $c_\pm(\omega)$ will appear again in the next section, where we
shall obtain their finite temperature version. The noise spectrum
to order $|\Gamma|^2$ is then given by
\begin{eqnarray}
S(\omega)&=&\frac{2\pi}{\Gamma(2g)} {e^*}^2|\Gamma|^2 \ [\
|\omega-\omega_0|^{2g-1}+|\omega+\omega_0|^{2g-1}\ ]
\nonumber \\
&=&e^*I_t\  [\ |1-\omega/\omega_0|^{2g-1}+|1+\omega/\omega_0|^{2g-1}\ ]\ ,
\end{eqnarray}
where we used the perturbative result to order $|\Gamma|^2$ for the
tunneling current $I_t=\frac{2\pi}{\Gamma(2g)}
e^*|\Gamma|^2\omega_0^{2g-1}$ \cite{XGW2}.

{}From the expression for $S(\omega)$ above we can deduce some features
of the noise to order $|\Gamma|^2$. First notice that for
$\omega\ll\omega_0$ we obtain $S(\omega)\approx 2e^*I_t$, the
classical shot noise result, independent of $g$. Notice also the
singularities at $\omega=\pm\omega_0$. In the particular case of
non-interacting electrons ($g=1$) we have $S(\omega)=2e^*I_t$ for
$|\omega|<\omega_0$, and $S(\omega)=2e^*I_t |\omega|/\omega_0$ for
$|\omega|>\omega_0$, which agrees, to lowest order in the transmission
coefficient $D$ (lowest order in $|\Gamma|^2$), with previous results
for the noise spectrum of non-interacting electrons \cite{Yang}. To
get the term in $D^2$ we need to go beyond this zeroth order
perturbation theory, as we will do later in the paper. The sharp edge
of the noise spectrum at $\omega=e^*V/\hbar$ for $g=1$ finds its
origin in the Pauli principle, which is the sole factor responsible
for correlations in the non-interacting case
\cite{Levitov&Lesovik}. In our model, particle statistics enter
automatically in the way we construct the electron/quasiparticle
operator from the boson fields and their commutation relations.

In the following sections we shall see how the low-frequency noise spectrum
is modified once we go beyond this perturbative calculation.

\section{The Joint Probability Distribution}

As we have previously mentioned, when the tunneling barrier is narrow
so that the time the charge spends in the tunneling process is small
compared to the times between two consecutive tunnelings, one can
speak of well defined tunneling events at certain specific times.  In
this section we will find a joint probability distribution for the
times for these tunneling events.

The term $e^{i\gamma\phi}$ in the Hamiltonian (where we use
$\gamma=\sqrt{g}$) transfers charge from one edge branch to the other
(say, in the case of the geometry of Fig. 1b \& 1c, from the $R$ to
the $L$ edge branch). The term $e^{-i\gamma\phi}$ does the converse (from
$L$ to $R$).  We will map the problem to a Coulomb gas in a 1-D space,
associating a charge $+$ to the term $e^{i\gamma\phi}$ and a charge
$-$ to $e^{-i\gamma\phi}$. Let $Z=\langle 0 | S_c(-\infty,-\infty) | 0
\rangle$, where $S_c(-\infty,-\infty)$ is the scattering operator in
the contour from $t=-\infty$ to $t=\infty$, and back to $t=-\infty$
(the Keldysh formalism contour). In terms of the usual scattering
operator $S$, we can write:
\begin{eqnarray}
Z&=&\langle 0|S(-\infty,\infty)\ S(\infty,-\infty)|0\rangle \nonumber\\
&=&\langle 0|S^\dagger (\infty,-\infty)\ S(\infty,-\infty)|0\rangle \ .
\end{eqnarray}
In this form the contributions from the forward
($t=-\infty\rightarrow\infty$) and return ($t=\infty\rightarrow
-\infty$) branches are easily identified in terms of the more commonly
used (equilibrium) scattering operators. Clearly, since $S$ is
unitary, $Z=1$. Now let's expand $Z$ perturbatively in $\Gamma$. We
will use the scripts $t$ and $b$ to denote the top (or forward) and
bottom (or return) branches, and $+$ and $-$ to denote whether the
inserted operator is $e^{i\gamma\phi}$ ($+$) or $e^{-i\gamma\phi}$
($-$).  $Q^{t,b}_{+,-}$ will denote the number of times that
$e^{i\gamma\phi}$ or $e^{-i\gamma\phi}$ appear in the top and bottom
contours (see Fig. 2). With this notation, we can expand the scattering
operator as:
\begin{eqnarray}
S(\infty,-\infty)=\sum_{Q^t_+,Q^t_-}\frac{(-i\Gamma)^{Q^t_+}\
(-i\Gamma^*)^{Q^t_-}}{Q^t_+!\ Q^t_-!} \int \prod_{i=1}^{Q^t_+} dt^{t+}_i\
\prod_{j=1}^{Q^t_-} dt^{t-}_j
\ T\ (\prod_{i=1}^{Q^t_+} e^{-i\omega_0 t^{t+}_i} e^{i\gamma\phi(t^{t+}_i)}\
\prod_{j=1}^{Q^t_-} e^{i\omega_0 t^{t-}_j} e^{-i\gamma\phi(t^{t-}_j)})\nonumber
\end{eqnarray}
and
\begin{eqnarray}
S(-\infty,\infty)&=&S^\dagger(\infty,-\infty)\nonumber\\
&=&\sum_{Q^b_+,Q^b_-}\frac{(i\Gamma)^{Q^b_+}\ (i\Gamma^*)^{Q^b_-}}{Q^b_+!\
Q^b_-!} \int \prod_{k=1}^{Q^b_+} dt^{b+}_k\ \prod_{l=1}^{Q^b_-} dt^{b-}_l
\ \tilde{T}\ (\prod_{k=1}^{Q^b_+} e^{-i\omega_0 t^{b+}_k}
e^{i\gamma\phi(t^{b+}_k)}\ \prod_{l=1}^{Q^b_-} e^{i\omega_0 t^{b-}_l}
e^{-i\gamma\phi(t^{b-}_j)})\nonumber
\end{eqnarray}
where $\tilde{T}$ stands for reverse time ordering. Notice that in the
operator $S_c(-\infty,-\infty)=S(-\infty,\infty)\ S(\infty,-\infty)$
the $\tilde{T}$ ordering occurs to the left of the $T$ ordering, so that we
replace both by a $T_c$ ordering operator such that times in the top
branch are ordered increasingly, times in the bottom branch are
ordered decreasingly, and times in the bottom branch are always
ordered after the ones in the top branch (see Fig. 2). Using $T_c$ we
can write $S_c(-\infty,-\infty)$ as

\begin{eqnarray}
&\ &\sum_{Q^t_+,Q^t_-,Q^b_+,Q^b_-}\frac{(-i\Gamma)^{Q^t_+}\
(-i\Gamma^*)^{Q^t_-}\ (i\Gamma)^{Q^b_+}\ (i\Gamma^*)^{Q^b_-}}{Q^t_+!\
Q^t_-!\ Q^b_+!\ Q^b_-!} \int \prod_{i=1}^{Q^t_+} dt^{t+}_i\
\prod_{j=1}^{Q^t_-} dt^{t-}_j \prod_{k=1}^{Q^b_+} dt^{b+}_k\
\prod_{l=1}^{Q^b_-} dt^{b-}_l\nonumber\\ &\ &\ \ \ \ \ \ \ \ \ \ \ \ \
\ \ \ T_c\ (\prod_{i=1}^{Q^t_+}\ e^{-i\omega_0 t^{t+}_i}
e^{i\gamma\phi(t^{t+}_i)}\
\prod_{j=1}^{Q^t_-} e^{i\omega_0 t^{t-}_j} e^{-i\gamma\phi(t^{t-}_j)}
\prod_{k=1}^{Q^b_+}\
e^{-i\omega_0 t^{b+}_k} e^{i\gamma\phi(t^{b+}_k)}\ \prod_{l=1}^{Q^b_-}
e^{i\omega_0 t^{b-}_l} e^{-i\gamma\phi(t^{b-}_l)})\ .\label{S_c}
\end{eqnarray}
In order to calculate the bracket
\begin{equation}
\langle 0|T_c (e^{i\gamma [\ \sum_{i=1}^{Q^t_+} \phi(t^{t+}_i)
-\sum_{j=1}^{Q^t_-} \phi(t^{t-}_j) +\sum_{k=1}^{Q^b_+} \phi(t^{b+}_k)
-\sum_{l=1}^{Q^b_-} \phi(t^{b-}_l)\ ]})|0\rangle\label{bracket}
\end{equation}
we use
\begin{equation}
\langle 0|T_c (e^{i q\phi(t)}\ e^{i q'\phi(t')})|0 \rangle=e^{-qq'\langle
0|T_c(\phi(t)\phi(t'))|0\rangle}
\end{equation}
and the contour-ordered two-point correlation
\[ \langle 0|T_c(\phi(t_1)\phi(t_2))|0\rangle=\left\{ \begin{array}{ll}
-2\ln (\delta +i|t_1-t_2|)&\mbox{, both $t_1$ and $t_2$ in the top branch}\\
-2\ln (\delta -i|t_1-t_2|)&\mbox{, both $t_1$  and $t_2$ in the bottom
branch}\\
-2\ln (\delta -i(t_1-t_2))&\mbox{, $t_1$  in the top and $t_2$ in the bottom
branch}\\
-2\ln (\delta +i(t_1-t_2))&\mbox{, $t_1$  in the bottom and $t_2$ in the top
branch}
\end{array}
\right. \]
\\

The bracket in Eq. (\ref{bracket}) contains the contributions from all
pairs of charges, which interact via a two body potential that is
determined by the $T_c$ ordered two-point correlation. The phase terms
due to $\omega_0$ ($e^{-i\omega_0 t}$ for a $+$ charge, and
$e^{i\omega_0 t}$ for a $-$ charge) correspond to an underlying
background, which tends to polarize the gas, leaving (in the case of
positive $\omega_0$, for example) more $+$ charges than $-$ ones in
the top branch, and more $-$ charges than $+$ ones in the bottom
branch.  An illustrative picture of the unbalance created by the
applied voltage $V$ (non-equilibrium) is shown in Fig. 3. One can
think of $V$ as an ``electric field'' that polarizes the
Coulomb gas, leaving an unbalance of $+$ and $-$ charges in the $t$
and $b$ contours, which gives rise to a net current in one direction
or the other (excess of $+$($-$) charges, or $R\rightarrow L$
($L\rightarrow R$) tunneling), depending on the sign of $V$.

The expression for $Z$ obtained as an expansion in $\Gamma$ is exact
so far. Also, the map into a Coulomb gas model is now complete.  An
expansion similar to the one we present here appears in the study of
dissipative quantum mechanics models in a periodic potential
\cite{Legget,Fisher&Zwerger}. There the charges are grouped in terms of the
so called ``soujourns'' and ``blips''. We find the idea of keeping the
$+$ and $-$ charges more intuitive, as is the idea of having the
non-equilibrium voltage be thought of as a ``field'' that polarizes
the gas and changes the densities within the $t$ and $b$ contours.
This language, as we will show, makes it easier for us to go beyond
the independent blip approximation, and study correlations.

We will now focus in showing how the expression for $Z$ can be used to
define a joint probability of tunneling events. In the limit of a
narrow barrier, as we pointed out previously, one can speak of rather
well defined tunneling times or tunneling events. In this limit we
can interpret the times that enter in the perturbative expansion of
$Z$ as the times for real tunneling events, and the sums and
integrations as the means of including all tunneling histories in a
partition function. Notice that it is very important that we
understand that this interpretation has a meaning only when the
tunneling barrier is narrow.

Also notice that only the tunneling times in the forward or top branch
can have a physical interpretation as a tunneling of a real charge (we
only observe increasing times, with the return branch being just a
mathematical tool). The correct joint probability distribution of
tunneling events should be obtained by integrating out all
$t^{b\pm}$'s. This is a difficult task, and we shall appeal to a more
intuitive picture that will allow us to sort out the most important
contributions. This more intuitive picture can be extracted from the
Coulomb gas model depicted in Figure 4.

The first step we take is to recast the sum in terms of dipole
configurations, as opposed to a sum of charge configurations. The
dipole is determined by a center of mass coordinate $t_{cm}$ and a
dipole strength $p$. There are four types of dipoles, as shown in Fig.
5. The type of dipole depends on which branches the $+$ and $-$
charges are located at.  We call a $t$ dipole one in which both
charges are in the top branch. A $b$ dipole is one where the charges
are in the bottom branch. In a $c_+$ the $+$ charge is on the top  and
the $-$ on the bottom. In a $c_-$ the converse is true, the $-$ is on
the top and the $+$ on the bottom. This distinction is important, as
we will see it later.

For a given charge configuration labeled by
$\{Q^t_+,Q^t_-,Q^b_+,Q^b_-\}$ we associate a dipole configuration
$\{n_t,n_b,n_+,n_-\}$, where the $n$'s are, respectively, the number
of $t$, $b$, $c_+$, and $c_-$ dipoles. The $n$'s and $Q$'s are related
by:
\begin{eqnarray}
Q^t_+=n_t+n_+ \ \ \ \ \ &\ & Q^t_-=n_t+n_-\nonumber\\
Q^b_+=n_b+n_- \ \ \ \ \ &\ & Q^b_-=n_b+n_+\ . \nonumber
\end{eqnarray}
Rewriting $Z$ in terms of the $n$'s instead of the $Q$'s becomes a
simple combinatoric task, which gives:
\begin{eqnarray}
Z&=&\sum_{n_t,n_b,n_+,n_-}\frac{(-i)^{Q^t_++Q^t_-}\
(i)^{Q^b_++Q^b_-}}{Q^t_+!Q^t_-!\ Q^b_+!\ Q^b_-!}\
|\Gamma|^{Q^t_++Q^t_-+Q^b_++Q^b_-}\nonumber\\
&\ &\nonumber\\
&\ &\ \ \ \ \times
\left( \begin{array}{c} Q^t_+\\n_t\end{array}\right)
\left( \begin{array}{c} Q^t_-\\n_t\end{array}\right)
\left( \begin{array}{c} Q^b_+\\n_b\end{array}\right)
\left( \begin{array}{c} Q^b_-\\n_b\end{array}\right)
n_t!n_b!n_+!n_-!
\times\  {\rm INTEGRAL}\nonumber\\
&\ &\nonumber\\
&\ &\nonumber\\
&=&\sum_{n_t,n_b,n_+,n_-}\frac{(-1)^{n_t+n_b}\ |\Gamma|^{2(n_t+n_b+n_++n_-)}}
{n_t!\ n_b!\ n_+!\ n_-!}\times\  {\rm INTEGRAL} \label{Zn's}
\end{eqnarray}
where the INTEGRAL term contains the interactions between the charges
integrated over all positions. The first approximation we will
make is what we will call the ``independent dipole'' approximation.
The attraction between opposite charges tends to bind them together,
and, if the fugacity of the gas (measured by $|\Gamma|^2$) is small,
we can to lowest order neglect the interaction between dipoles. The only
interactions entering in the calculation of $Z$ are the intra-dipole
interactions. The INTEGRAL term in the dipole approximation can be
factored as a product of the contributions of individual dipoles.
\begin{equation}
{\rm INTEGRAL}=t^{n_t}\ b^{n_b}\ c_+^{n_+}\ c_-^{n_-}
\end{equation}
where
\begin{equation}
t=\int_{-\infty}^\infty dp \frac{e^{-i\omega_0 p}}{(\delta + i|p|)^{2g}}\ ,
\ b=\int_{-\infty}^\infty dp \frac{e^{-i\omega_0 p}}{(\delta - i|p|)^{2g}}\ ,
\ c_\pm=\int_{-\infty}^\infty dp \frac{e^{-i\omega_0 p}}{(\delta \mp ip)^{2g}}\
{}.
\end{equation}
One can check that $t+b=c_++c_-$, so that summing over all $n_t$ and
$n_b$ in Eq. (\ref{Zn's}) can be shown to yield:
\begin{equation}
Z=e^{-|\Gamma |^2(c_++c_-)}\ \sum_{n_+,n_-}\ \frac{(|\Gamma |^2
c_+)^{n_+}}{n_+!}
\frac{(|\Gamma |^2 c_-)^{n_-}}{n_-!}\ . \label{Zdipole}
\end{equation}
Let us now interpret this expression. As we mentioned above, only
events occurring in the forward or top branches can be observed.
Therefore, the occurrence of a dipole of the $c_+$ type implies a
tunneling event in one direction occurring at the vicinity of the
center of mass coordinate of the dipole.  Conversely, a dipole $c_-$
implies a tunneling event in the opposite direction. The statistical
distribution of these center of mass coordinates of dipoles appears in
the noise. The uncertainty of the location of the charges comprising
the dipole with respect to the dipole center of mass also contributes
to the noise; this intra-dipole contribution, however, is already
partly taken care of in the first order perturbative calculation of
noise, which can be seen to be nothing but the correlation between the
position of the two charge components of a dipole.  The intra-dipole
noise is in the high-frequency range, centered at
$\omega=\omega_0=e^*V/\hbar$. The contribution to the noise that we
obtain with the $Z$ in Eq.(\ref{Zdipole}) is in the low-frequency
range ($\omega\ll\omega_0$), where the positions of the charges and
dipole centers are not distinguished.  The reason why we summed over
the dipoles of type $t$ and $b$ is that they do not contribute to the
noise beyond the intra-dipole order. These types of dipole correspond
to tunneling in one direction shortly followed by tunneling in the
opposite direction, which contribute to noise in the time scale of the
dipole size, included in the intra-dipole contribution.

With the interpretation above in hand, we can use Eq. (\ref{Zdipole})
to argue that, in the dipole approximation, the tunneling events in
either direction are independent, with a distribution that is
Poisson-like with two parameters: $|\Gamma |^2 c_+$ and $|\Gamma |^2
c_-$. The probability of tunneling in one direction in an infinitesimal
time $\Delta t$ is $P_+=|\Gamma |^2 c_+\ \Delta t$, the probability of
tunneling in the opposite direction is $P_-=|\Gamma |^2 c_-\ \Delta
t$, and the probability of no tunneling event in this time is
$1-(P_++P_-)$.

This two-parameter Poisson distribution can be used to reproduce the
results obtained for the tunneling current to first order in
perturbation theory.  The tunneling current is simply given by
$I_t=e^*|\Gamma |^2 (c_+-c_-)$, {\it i.e.}, the net rate of tunneling
in one direction. To obtain an expression for $I_t$ in terms of $V$ we
need to evaluate $c_+$ and $c_-$:
\begin{equation}
c_+(\omega_0)=c_-(-\omega_0)=\int_{-\infty}^\infty dp \frac{e^{-i\omega_0
p}}{(\delta - ip)^{2g}}=
2\pi\frac{\ \ \ |\omega_0|^{2g-1}}{\Gamma(2g)}\ \theta(\omega_0)\ .\label{cT=0}
\end{equation}
We can obtain the finite temperature results for $c_\pm$ by a
conformal transformation \cite{Shankar}:
\begin{equation}
\int_{-\infty}^\infty dp \frac{e^{-i\omega_0 p}}{(\delta - ip)^{2g}}\rightarrow
e^{i\pi g}\int_{-\infty}^\infty dp
\frac{e^{-i\omega_0 p}}{|\frac{\sinh(\pi Tp)}{\pi T}|^{2g}}\ ,
\end{equation}
which gives
\begin{equation}
c_+(\omega_0)=c_-(-\omega_0)=2(\pi T)^{2g-1}B(g+\frac{i\omega_0}{2\pi
T},g-\frac{i\omega_0}{2\pi T})
\ \cosh(\frac{\omega_0}{2T})\ (1+\tanh(\frac{\omega_0}{2T}))\ ,\label{c's}
\end{equation}
where $B$ is the Beta function. Using these expressions for $c_\pm$ we obtain
\begin{equation}
I_t=4e^*|\Gamma |^2 (\pi T)^{2g-1}B(g+\frac{i\omega_0}{2\pi
T},g-\frac{i\omega_0}{2\pi T})
\ \sinh(\frac{\omega_0}{2T})\ ,\label{I_t}
\end{equation}
which is the same expression found by first order perturbation theory
in Ref. \cite{XGW2}. For $T=0$, in particular, we find that $I_t\sim
e^*|\Gamma |^2 V^{2g-1}$.

We now turn to the noise properties derived from this dipole
approximation. Because the distribution in this approximation is
Poisson like (and therefore uncorrelated), we should expect the noise
to have a flat frequency dependence, {\it i.e.}, white noise. We are
left with the problem of determining the amplitude of the noise. For
this purpose, we will follow a calculation similar to one presented by
Landauer \cite{Landauer1}. Let $\langle j^2\rangle_{\Delta f}$ be the
component of the noise power spectrum that falls in the frequency
interval $\Delta f$. Let also $S(f)=\int_0^\Theta dt\ j(t)\
e^{-i\omega_0 t}$, where $\Theta$ is a time interval. These quantities
are related by:
\begin{equation} \langle j^2\rangle_{\Delta
f}=\lim_{\Theta\rightarrow\infty}\frac{2|S(f)|^2}{\Theta}\ \Delta f\ .
\end{equation}
The charge transferred in a small interval of time $\tau$ is $\pm e^*$
(with probabilities $|\Gamma |^2 c_\pm\tau$), or 0. We can write
$j(t)=\sum_n j_0(t-n\tau)\ q_n$, with $q_n=\pm1,0$. Here $j_0$ is a
narrow current pulse that fits a slot of time $\tau$ (the width of the
pulse should determine a cut-off frequency above which the spectrum is
no longer flat). We can then write
\begin{equation}
S(f)=\int_0^\Theta dt e^{-i\omega t} \sum_n j_0(t-n\tau)
=\sum_n e^{-i\omega n\tau} q_n \int_{-n\tau}^{\Theta -n\tau} du\ e^{-i\omega u}
j_0(u)\ .
\end{equation}
The last integral can be approximated by the total charge that tunnels
($e^*$), since the pulse is narrow compared to $\tau$. We then have
$S(f)=e^* \sum_n e^{-i\omega n\tau} q_n$, and thus
\begin{equation}
|S(f)|^2={e^*}^2 \sum_{n,n'} e^{-i\omega n\tau} e^{i\omega n'\tau} q_n q_{n'}\
{}.
\end{equation}
The uncorrelated tunneling implies that $\langle q_n q_{n'}\rangle=\langle
q\rangle^2+(\langle q^2\rangle-\langle q\rangle^2)\delta_{n,n'}$.
After summing over $n$ and $n'$ we obtain that $|S(f)|^2={e^*}^2 N
(\langle q^2\rangle-\langle q\rangle^2)$, where $N=\Theta/\tau$ is the
number of time slots. Now $\langle q\rangle=|\Gamma |^2
(c_+-c_-)\ \tau$ and $\langle q^2\rangle=|\Gamma |^2 (c_++c_-)\ \tau$, and
for small tunneling times compared to the time between tunneling
$\langle q\rangle\ll 1$, so that we can neglect $\langle q\rangle^2$
and obtain
\begin{equation}
\langle j^2\rangle_{\Delta f}=2{e^*}^2|\Gamma |^2 (c_++c_-)\Delta f\ .
\end{equation}
We can connect the white noise amplitude to the tunneling current
using Eqs. (\ref{c's}) and (\ref{I_t}), and obtain
\begin{equation}
\langle j^2\rangle_{\Delta f}=2e^*I_t\ \coth(\frac{\omega_0}{2T})\Delta f\ .
\end{equation}
If we write $I_t=GV=G\omega_0/e^*$ and take the $\omega_0\rightarrow 0$
limit, we obtain $\langle j^2\rangle_{\Delta f}^{eq.}=4TG\Delta f$,
which is nothing but the Johnson-Nyquist equilibrium ($V=0$) result.
The non-equilibrium white noise can then be cast in a simple relation
to the equilibrium Johnson-Nyquist noise, which is
\begin{equation}
\langle
j^2\rangle_{\Delta f}=(\frac{e^*V}{2T})\coth(\frac{e^*V}{2T})\ \langle
j^2\rangle_{\Delta f}^{eq.}\ .
\end{equation}
The expression above for $T\rightarrow 0$ gives $\langle
j^2\rangle_{\Delta f}=2e^*I_t\ \Delta f$, which is the classical
expression for shot noise. Quantum corrections to the shot noise only
come to order $|\Gamma|^4$ and higher, and thus do not appear in the
independent dipole approximation (order $|\Gamma|^2$).
Also notice that the expression connecting equilibrium and
non-equilibrium noise $\frac{e^*V}{2T}\coth(\frac{e^*V}{2T})$ is
universal independent of $g$ and thus independent of interactions.
This is consistent with the fact that the independent dipole
approximation is a lowest order perturbative result, so that the
assumptions necessary for the fluctuation-dissipation theorem are
satisfied.

The dipole approximation therefore captures the uncorrelated part of
the noise. In the next section we shall see how correlations come
about.

\section{Beyond the Independent Dipole Approximation}

In this section we shall improve the dipole approximation. We have
seen that the location of the centers of mass of two dipoles is
uncorrelated in the approximation of the preceding section. In order
to observe correlations one must include in the model the interactions
between distinct dipoles. This is the next order correction to the
INTEGRAL term in Eq. (\ref{Zn's}).

Consider two dipoles as show in Fig. 6. We take them, for the sake of
illustration, to be both of the $c_+$ type. The INTEGRAL term can be
written for this case as:
\begin{eqnarray}
&\int& dt_1dt_2dp_1dp_2 \frac{e^{-i\omega_0 p_1}}{(\delta -ip_1)^{2g}}
\frac{e^{-i\omega_0 p_2}}{(\delta -ip_2)^{2g}}\\
&\ &\nonumber\\
&\ &\ \ \ \ \ \times\
\frac{[\delta +i(t_2+p_2/2-t_1-p_1/2)]^{2g}\ [\delta
-i(t_2-p_2/2-t_1+p_1/2)]^{2g}}
{[\delta +i(t_2-p_2/2-t_1-p_1/2)]^{2g}\ [\delta
-i(t_2+p_2/2-t_1+p_1/2)]^{2g}}\nonumber\ .
\end{eqnarray}
For dipole separations that are large compared to dipole sizes
($|t_2-t_1|\gg |p_1|,|p_2|$), we can expand the expression in the
integrand to obtain
\begin{equation}
\int dt_1dt_2dp_1dp_2 \frac{e^{-i\omega_0 p_1}}{(\delta -ip_1)^{2g}}
\frac{e^{-i\omega_0 p_2}}{(\delta -ip_2)^{2g}}
\left[1+2g\frac{p_1p_2}{(t_2-t_1)^2}\right]\ ,
\end{equation}
which after the $p_1$ and $p_2$ integration yields
\begin{equation}
\int dt_1dt_2 \left(c_+c_+\ -\ 2g\frac{c'_+c'_+}{(t_2-t_1)^2}\right)
\approx c_+c_+ \int dt_1dt_2 \
e^{-2g\frac{(c'_+/c_+)(c'_+/c_+)}{(t_2-t_1)^2}}\ .
\end{equation}
This can be generalized to any two types of dipole to
\begin{equation}
d_1d_2 \int dt_1dt_2 \  e^{-2g\frac{(d'_1/d_1)(d'_2/d_2)}{(t_2-t_1)^2}}\
,\label{dipole-dipole}
\end{equation}
where $d_{1,2}$ can be any of $t$, $b$, $c_+$ or $c_-$, and $d'_{1,2}$
stands for the derivative of $d_{1,2}$ with respect to $\omega_0$.
Using a similar argument to the one we used to obtain the finite
temperature expression for $c_\pm(\omega_0)$, we can obtain the finite
temperature version of the dipole-dipole interaction by simply
substituting $t_2-t_1$ by $\sinh[\pi T (t_2-t_1)]/(\pi T)$ and using
the $T\neq 0$ results for $c_\pm(\omega_0)$. Nevertheless, we will
just concentrate for the rest of the paper on the $T=0$ problem.

{}From Eq. (\ref{dipole-dipole}) we read that the dipoles interact
through a $1/t^2$ potential. This dipole-dipole interaction gives rise
to a non-trivial distribution of tunneling events, which show up in
the noise spectrum as a cusp at zero frequency. Before proceeding to
obtain the explicit form, including the strength of the singularity,
we must understand when this picture that the charges can be assembled
in pairs starts to break down, and correlations not contained in this
dipole picture become important.

The assumption we made in order to obtain correlations as in Eq.
(\ref{dipole-dipole}) was simply that the dipole sizes were small
compared to the separation between dipoles. The mean dipole separation
is related to the average current $I_t$, and is given by $e^*/I_t$.
The dipole size can be taken to be the $d'/d$ in Eq.
(\ref{dipole-dipole}), since it is this term that enters in the
interaction between the dipoles, and thus measures the distance
between the $+$ and $-$ charges that form the dipole (Notice that,
because the charges in the Coulomb gas are $\pm 1$, the distance
between the $+$ and $-$ charges equals the dipole strength). The
expressions for $t$ and $b$ depend on the cut-off scale $\delta$,
whereas $c_\pm$ are finite as we take $\delta\rightarrow 0$ (we can
show that $c_++c_-=t+b$, and that the divergences in $t$ and $b$,
which are purely imaginary, cancel each other). We then have that
$t'/t$ and $b'/b$ must both scale as $\delta$, and
$c'_\pm/c_\pm=(2g-1)\omega_0^{-1}$ (using Eq. (\ref{c's}) and setting
$T\rightarrow 0$). Therefore the dipole approximation is good as long
as $\omega_0^{-1}\ll e^*/I_t$, or $I_t\ll ({e^*}^2/h)V$.

In the case of tunneling between edge states, this is the limit of a
small tunneling current as compared to the Hall current. In the case
of the 1-D channel, this limit corresponds to a small tunneling
current as compared to the current for the non-interacting case.
Because of the non-linear $I-V$ characteristic of the tunneling
current in 1-D Luttinger liquids ($I_t\propto V^{2g-1}$), the cases
$g>1$ and $g<1$ are quite distinct. For $g>1$ the dipole phase exists
for small applied voltages $V$, whereas for $g<1$ the dipole phase
exists for large $V$. Now, one can still use the results of the dipole
phase to study the noise in the case of $g>1$ and large $V$, and the
case of $g<1$ and small $V$, by resorting to the duality $g
\leftrightarrow 1/g$ that connects the two configurations shown in
Figs. 1b \& 1c. The idea is that as one increases the applied voltage
between the $R$ and $L$ edges in the configuration shown in Fig. 1c,
the tunneling current $I_t$ increases asymptotically, tending to the
Hall current. Deviations from the Hall current correspond to
``defects'', or tunneling in the direction perpendicular to the Hall
current, which is exactly the direction of tunneling shown in Fig. 1b.
Similarly, one can go from the situation in Fig. 1b to the one in Fig.
1c by decreasing the applied voltage between the $R$ and $L$ edges.
The bottom line is that, by wisely choosing which current direction to
focus on, one can most often place the problem in the dipole limit for
either one configuration or its dual with $g \leftrightarrow 1/g$. The
regime in which the dipole picture fails is then at the crossover
between the two configurations, where the gas will be in a plasma
phase.

Now that we understand when the approximation is valid, let us look at
its consequence in the noise spectrum. At zero temperature we only have
either one of $c_+$ or $c_-$ dipole types, depending on the sign of $\omega_0$
(see Eq. (\ref{cT=0}) ). For concreteness, let us take $\omega_0 >0$,
so that $c_+$ dipoles survive. Since $t'/t,b'/b\sim \delta$, the main
correlations come from the interactions between $c_+$ dipoles (for voltages
small compared to $1/\delta$), so that for large times the density-density
correlation for $c_+$ dipoles (which equals the current-current correlation)
is given by
\begin{equation}
\langle\rho_+(t)\ \rho_+(0)\rangle_c\sim \langle\rho_+\rangle^2
\frac{-2g(2g-1)^2}{\omega_0^2 t^2},
\end{equation}
which gives a noise spectrum $S(\omega)= 4\pi g(2g-1)^2
\left(\frac{I_t}{\omega_0}\right)^2 e^{-|\omega|\lambda_c}/\lambda_c$, where
$\lambda_c$ is a short time scale cut-off (of order $\omega_0^{-1}$)
for the $1/t^2$ correlation.  The leading singularity at low-frequency
is then
\begin{equation}
S_{sing.}(\omega)=4\pi g(2g-1)^2 (\frac{I_t}{\omega_0})^2\ |\omega|\ .
\end{equation}
Since $I_t\propto V^{2g-1}$, the strength of the singularity has a
non-linear dependence $V^{4(g-1)}$ on the applied voltage.

For the particular case of non-interacting electrons ($g=1$) one can
write $\frac{I_t}{\omega_0}=\frac{e^*D}{2\pi}$, where $D$ is the
transmission coefficient. The noise spectrum singularity is then
$S_{sing.}(\omega)=\frac{{e^*}^2 D^2}{\pi} |\omega|$, which recovers
the result of Ref. \cite{Yang}. The effects of correlation due to the
Pauli principle enter automatically in our formulation of the problem
through the bosonization.

To finish this section, let us consider the case of equilibrium noise
within the interacting dipole approximation. For $g>1$ the tunneling
current vanishes for $V=0$. In the case of $g<1$, however, we have to
invoke the dual picture ($g \rightarrow 1/g$) in order to use the
dipole language. In any case the $|\omega|$ singularity due to the
dipole-dipole interaction vanishes for $V=0$. The reason can be viewed
very simply: the non-equilibrium voltage was responsible for the
polarization of the dipole gas, and the dipole-dipole interaction gave
the $1/t^2$ correlation. At equilibrium, the average dipole strength
vanishes, and the interactions in this case must come from induced
dipole, or ``Van der Waal's'' attraction, which for our log potentials
goes as $1/t^4$. We can show that the low-frequency behavior of the
noise no longer has the $|\omega|$ singularity, but has leading
contributions from $\omega^2$ and $|\omega|^3$. The leading
singularity is then $\propto |\omega|^3$. The contributions calculated
above are only the inter-dipole correlations. We should also account
for the intra-dipole correlations, because for $V=0$ the singularity
at the ``Josephson'' frequency falls to $\omega=0$. We already calculated
the intra-dipole contribution to the noise spectrum perturbatively in
section III, which for zero applied voltage is $S(\omega)\propto
|\omega|^{2g-1}$. At low frequencies, the intra-dipole noise will be
dominant for $g<2$, while the inter-dipole noise will be dominant for
$g>2$. Notice that, in the equilibrium case, an expansion for $Z$ like
the one in Eq. (\ref{S_c}) could be carried out with only one branch,
since in equilibrium there is no need for the two branches of the
Keldysh contour.

\section{Diagrammatic Technique}
The dipole gas picture we used to expand $Z$ can be justified in more
formal manner. In this section we shall present a systematic way to
expand $Z$ diagrammatically, which is used in one-dimensional
dissipative quantum mechanics models \cite{DEF}. In this expansion we
can identify the terms we included in the dipole picture. The
expansion is the formal support for the more intuitive and physical
picture of the dipole gas. We will first present an introduction to
the diagrammatic expansion, followed by the calculation for the
equilibrium case and implications for the non-equilibrium case.

\subsection{Introduction to the Diagrammatic Expansion}

We start by returning to the expansion of $S(-\infty,-\infty)$ in
terms of the bare charges in Eq. (\ref{S_c}). We will focus on the
expectation value of the $T_c$ ordered product. Let us use a slightly
different notation, using $t$'s to denote the positions of $+$ charges
and $s$'s to denote the positions of $-$ charges. Let us take some
configuration of charges labeled by $t_i$ and $s_j$, with some of them
on the top and some on the bottom branch (this way we do not have to
worry about the superscripts for top and bottom branches, since we can
keep track of where each charge is by its index).
Using this notation, we can write for the $T_c$ bracket:
\begin{equation}
P = \left(\frac{\prod_{i<j\le Q}[\delta +i(t_i-t_j)\alpha_c(t_i,t_j)]
\prod_{i<j\le Q}[\delta +i(s_i-s_j)\alpha_c(s_i,s_j)]}
{\prod_{i,j}[\delta +i(t_i-s_j)\alpha_c(t_i,s_j)]} \right)^{2g}\ ,
\label{PKeldysh}
\end{equation}
where $Q=Q^t_++Q^b_+=Q^t_-+Q^b_-$, and $\alpha_c(t,t')=\pm 1$ depending on the
ordering of $t$ and $t'$ along the Keldysh contour.

Consider now integer $g$'s, such that $2g$ is even and the expression above
does not change if we take $[\delta +i(t-t')\alpha_c(t,t')]\rightarrow
[\delta \alpha_c(t,t') +i(t-t')]$. The expression for the $T_c$ bracket
can be simplified with the aid of the following identity which can be
proved using partial fractions or properties of determinants \cite{Muir}:
\begin{equation}
\frac{\prod_{i<j}(z_i-z_j)\prod_{i<j}(w_i-w_j)}{\prod_{i,j}(z_i-w_j)}
={\rm det}M(z,w)\ , \label{Midentity}
\end{equation}
where M is a matrix defined by
\begin{equation}
M_{ij}=\frac{1}{z_i-w_j}\ .
\end{equation}
The presence of the regulators $\delta$ in the expression for the $T_c$
bracket slightly complicates how we apply the identity to the problem.
By naively defining
\begin{equation}
M_{ij}=\frac{1}{\delta \alpha_c(t_i,s_j)+i(t_i-s_j)}
\end{equation}
we would obtain terms in the numerator to order $\delta$ and higher
that would not match the numerator of the expression for the $T_c$
bracket. This corresponds to a different choice of regularization, and
we shall return to this point later. The leading term (order
$\delta^0$), however, is exactly the same, and we proceed with the
program, writing $({\rm det}M)^{2g}$ for the $T_c$ bracket.

Now notice that the terms $M_{ij}$ correspond to the interaction between
$+$ and $-$ charges, and that the expansion of the determinant will be
comprised of all ways of combining $+$ to $-$ charges in pairs such that
each charge only appears once in the expansion. Let us then associate
a graph for any such dipole combination, as show in Fig. 7. When we
raise the determinant to the power $2g$, the effect is to obtain all
different ways to connect the $+$ and $-$ charges with lines, such
that each charge is connected by exactly $2g$ lines (see Fig. 7, where
we illustrate the case of $g=1$).

The graphs so obtained give us a systematic way to account for the
contributions to $Z$. The terms in the expansion where all charges are
connected to one and only one other charge, as in Fig. 8a, are the
independent dipole terms. Notice that each line in the graph that
connects two distant charges roughly corresponds to $1/t$.
When there are 4 charges, the lowest
order in $1/t$ that can be obtained from the expansion comes from taking
two dipoles and using one line from each to connect it to the other, so that
there are 2 lines connecting the dipoles (Fig. 8b). In this way
we obtain a $1/t^2$
term which corresponds to the leading dipole-dipole interaction.  This
systematic way to expand $Z$ can be used as the formal support for the
dipole picture developed in the previous section.

\subsection{Equilibrium Case}

To illustrate the power of the formalism described in the previous subsection,
we will consider the equilibrium case at zero temperature.  According to the
dipole approximation, we expect that the current-current correlation should
go as $1/t^4$ for $g \ge 2$.  In this section we will show this is the
case for any integer $g \ge 2$.

In equilibrium, we no longer need to use the Keldysh contour.
Instead, to simplify the calculations we will work in Euclidean space,
and we will take the $\phi$ field correlator to be $\langle 0 |
\phi(t) \phi(0)|0\rangle = - \ln(t^2 + \delta^2)$.  This differs from
the original choice of correlation function only in that we have used
a different cut-off.  The choice given here corresponds to left and
right movers scattering off each other instead of left movers with
left movers, as in the original choice.  However, for a single
impurity, left movers and right movers should really be equivalent, so
this choice should not make any important difference in the results.
More importantly, in both cases we choose to regulate the correlator
$\langle 0|\phi(t) \phi(0) |0\rangle$ consistently, no matter where it
appears in the expression for $P$ in Eq. (\ref{PKeldysh}).  It may
appear that whenever two tunneling events with the same charge
interact, we could just ignore the cut-off in the numerator of $P$,
since $\langle e^{i\gamma\phi(t)} e^{i\gamma\phi(0)}\rangle = (t^2 +
\delta ^2)^g$ is not singular as $\delta$ goes to zero; recall that
$\gamma = \sqrt{g}$.  However, the $\delta$'s in this correlator will
be multiplied by other correlators that are singular as $\delta$ goes
to zero, so it turns out that the answer depends on how we regulate
the numerator.  Because we are using the Coulomb gas picture, for now
we will choose to keep the $\delta$'s in the numerator.

With our choice of regulator, the expression $P$ for the bracket
needed to evaluate $S(-\infty, -\infty)$ becomes
\begin{equation}
P = \left({\prod_{i<j\le Q} ((t_i-t_j)^2 + \delta^2)
         \prod_{i<j\le Q} ((s_i-s_j)^2 + \delta^2) \over
         \prod_{i,j} ((t_i-s_j)^2 + \delta^2)} \right)^g.
\label{Pdef}
\end{equation}
In this equation, the positive charges are at the $t_i$ and the negative
charges are at the $s_i$.
Because we are in Euclidean space, we no longer have to use time-ordering
when we evaluate the integrals over the $t_i$ and $s_i$.

To simplify the expression for $P$ for any integer $g$,
we will use the same procedure as in
Ref. \cite{DEF}.  We will write $P = AB$, where $A$ equals $P$ with the
$\delta$'s in the numerator set to zero, and $B$ is the correction due to
the $\delta$'s in the numerator of $P$.  Then
\begin{equation}
A =     \left({\prod_{i<j\le Q} (t_i-t_j)^2
         \prod_{i<j\le Q} (s_i-s_j)^2 \over
         \prod_{i,j} ((t_i-s_j)^2 + \delta^2)} \right)^g,
\end{equation}
and $B$ is equal to sums over products of $\delta^2/(t_i-t_j)^2$ and
$\delta^2/(s_i-s_j)^2$, where any one of these expressions can occur at
most $g$ times in a product.  $B$ comes from writing each correlator in the
numerator as $(t_i-t_j)^2 \left[1 + \delta^2/(t_i-t_j)^2\right]$ and
factoring out the $(t_i - t_j)^2$ part.

We can again use the identity in Eq. (\ref{Midentity}) to simplify
the expression for $A$.  If we define the matrix $M(\delta)$ as
\begin{equation}
M_{ij}(\delta) = {1\over t_i - s_j + i\delta},
\end{equation}
then $A$ is given by
\begin{equation}
A = \left[{\rm det}M_{ij}(\delta) {\rm det}
M_{ij}(-\delta)\right]^g.
\label{AMM}
\end{equation}
As explained in the previous subsection, if we represent each charge by a point
and each factor of ${1\over t_i - s_j \pm i\delta}$ by a directed line, then
we obtain all the different ways to connect the positive charges to the
negative charges so that each charge is connected by exactly $2g$ lines,
(half of which are pointing toward the line, and half away from the line).

We can also give a graphical interpretation of B.  Once we have a graph from
A, to take into account the fact that the numerator is also regulated, we
obtain our graphs for P by joining any number of pairs of similar charges
with the pair of edges $1/(t_i-t_j)^2$ or $1/(s_i-s_j)^2$.  Each of these
edges is accompanied by a factor of $\delta$, and any pair of charges can
be joined by at most $g$ of these pairs of edges.  Thus $B$ introduces an
interaction between like-charged particles.

In the graphs of $A$ and $B$, it is important to keep track of the
number of vertices, $V$, the number of edges, $E$, and the number of
factors of $\delta$ in the numerator, $-f$.  If we are calculating the
charge-charge correlation function, and we insert $2N$ additional
charges, then the number of vertices is $V= 2N + 2$.  For any
connected graph of $A$, we then have $E = (2N + 2)g$ and $f = 0$.
Once we include the effects of $B$, $f$ is no longer equal to $0$, but
$E+f$ is still given by
\begin{equation}
E+f = (2N+2)g.
\end{equation}
Also, it is important to note that any connected graph is also 1PI.  This
way of describing the bracket, $P$, works similarly in Minkowski space.

Next, we will evaluate the connected correlation function of $\langle
0|e^{i\gamma\phi(t)}e^{-i\gamma\phi(s)}|0\rangle$ for any integer $g >
1$.  (The case when $g=1$ was considered in Ref. \cite{DEF}.)  This
calculation will also work for the correlation functions $\langle
0|e^{\pm i\gamma\phi(t)}e^{\pm i\gamma\phi(s)}|0\rangle$, so that
these results can be used to find the leading dependence on $t-s$ of
the current-current correlation functions.

At the $(2N)^{\rm th}$ order in perturbation theory, we have
\begin{equation}
\langle 0 | \Gamma e^{i\gamma\phi(t)}\Gamma^* e^{-i\gamma\phi(s)}|0\rangle
= \int_{-\infty}^\infty {|\Gamma|^{2N + 2}\over N! N!}
\prod_{k=1}^N dt_k \prod_{k=1}^N ds_k A B,
\end{equation}
where $A$ and $B$ depend on $t$, $s$, the $t_k$'s and the $s_k$'s.  To
obtain the connected correlation function, we just need to consider the
connected graphs in the expression on the right-hand side of the above
equation.

To evaluate the integrals, we will perform contour integrals where we
complete the contour in the upper half plane.  Thus, for each vertex,
$t_j$, we will be evaluating residues for all the poles occurring at
$t_j = s_k + i \delta$.  (Here, we are using $t_j$ and $s_k$ to stand
for any type of vertex.)  We note that in $B$, it appears that we will
have poles on the real axis.  However, we know that the original
expression for $AB$ does not have any poles on the real axis.  This
means that if we sum over all the graphs for $B$, these poles cancel,
which implies that as long as we integrate over the variables in each
of these graphs in the same order, we can just ignore the poles on the
real axis.

We can describe the process of evaluating residues diagrammatically,
as explained in Ref. \cite{DEF}.  If the multiplicity of the pole at
$t_j = s_k + i\delta$ is equal to one, then there is only one edge,
$e_{jk}$, that joins $t_j$ to $s_k$ and represents this kind of pole.
In this case, when we evaluate the residue, we just ``collapse" the
vertex $t_j$ and the edge $e_{jk}$.  This means we remove the vertex
$t_j$ and edge $e_{jk}$, and then reconnect all the other edges that
were originally connected to $t_j$ to $s_k$ instead.  If the other
end-point of any of these edges was also connected to $s_k$, the edge
becomes $1/(ic\delta)$, for some integer $c$.  Otherwise, it remains
an edge.  Consequently, the total number of edges decreases by at
least one, and the sum of edges and factors of $\delta$ in the
numerator decreases by exactly one.  Also, the graph remains connected
and 1PI.

When the multiplicity, $m$, of a pole is greater than one, then instead of
collapsing only one edge, we must collapse all the $m$ edges that
correspond to the pole.  In addition, we must take $m-1$ derivatives with
respect to $t_j$.   Each of these derivatives increases the number of other
legs connected to $t_j$ by one, so we obtain $m-1$ new legs.  Once we have
created these $m-1$ new edges and collapsed both the vertex and the $m$ edges
corresponding to the pole, we again find that the number of edges decreases
by at least one, and the sum, $E+f$, still decreases by exactly one.  Again,
the graph remains 1PI.

Now we can count the number of edges and factors of $\delta$ that
remain after we have done all the integrations.  The original graph
with $2N+2$ vertices has $E+f = (2N+2)g$.  After we integrate over the
$2N$ inserted charges, this sum becomes
\begin{equation}
E+f = (2N+2)g -2N,
\end{equation}
and the only two remaining vertices are $t$ and $s$.  Because the
graph must still be 1PI, we must have at least 2 edges connecting $t$
and $s$.  Since the total number of edges always decreases by at least
one, we also have
\begin{equation}
2 \le E \le (2N+2)g - 2N.
\end {equation}
We will let $l_N = (2N+2)g - 2N$.  Lastly, because the final answer
must be symmetric in $t$ and $s$, after we sum over all the graphs we
can only have even values for $E$.

Putting all of this together, we find that the correlation function of
$\langle 0 | \Gamma e^{i\gamma\phi(t)}\Gamma^* e^{-i\gamma\phi(s)}|0\rangle$
must have
the form
\begin{equation}
\Gamma \Gamma^* {1\over t^{2g}} + \sum_{N=1}^\infty
\left(\Gamma\Gamma^*\right)^{(N+1)}
\left(a_{N2}{1\over\delta^{l_N-2}(t-s)^2} +
a_{N4}{1\over\delta^{l_N-4}(t-s)^4} + \dots +
a_{l_N} {1\over(t-s)^{l_N}}\right),
\end{equation}
where the $a_{Nj}$'s are constants that are determined from
integrating the explicit graphs.  In order to interpret these results,
for the equilibrium case it is helpful to renormalize the coupling.
We will replace each $\Gamma$ and $\Gamma^*$ with $\Gamma\delta^{g-1}$
and $\Gamma^*\delta^{g-1}$.  This just takes into account the
self-interaction of the charges and a rescaling of the time variables.

The correlation function is then
\begin{equation}
\Gamma \Gamma^* {\delta^{2g-2}\over t^{2g}} + \sum_{N=1}^\infty
\left(\Gamma\Gamma^*\right)^{(N+1)}
\left(a_{N2}{1\over(t-s)^2} +
a_{N4}{\delta^2\over(t-s)^4} + \dots +
a_{l_N} {\delta^{l_N-2}\over(t-s)^{l_N}}\right).
\label{propsum}
\end{equation}
This general form is true to all orders in $\Gamma$.  Also, note that
the derivation of this result did not depend on the sign of the
charges at $t$ and $s$, so we will obtain a similar expression for two
positive charges or two negative charges at $t$ and $s$.  For large
times, (or small cut-off $\delta$) the leading behavior is
\begin{equation}
{1\over (t-s)^2 }\sum_{n=1}^\infty(\Gamma\Gamma^*)^{N+1}a_{N2}
+{\delta^2\over(t-s)^4}\sum_{N=1}^\infty(\Gamma\Gamma^*)^{N+1} a_{N4}.
\end{equation}
This expression appears to go as $1/(t-s)^2$ instead of as the
$1/(t-s)^4$ predicted by the dipole picture.  However, as we shall
show shortly, if both the denominator and numerator are regulated in
the same way, as in Eq.(\ref{Pdef}), then $a_{2N} = 0$ for all $N$, so
the leading behavior does go as $\delta^2/(t-s)^4$ to all orders in
perturbation theory.

Before showing that $a_{N2}=0$ for all $N$, we will first use the
previous calculation to describe a systematic way to determine the
leading behavior of each graph.  First, we note that a final answer of
$1/(t-s)^n$ corresponds to a graph with $n$ legs joining the vertex
$t$ to the vertex $s$.  If we remove these $n$ legs, the graph breaks
into 2 disconnected components, one containing $t$, and the other
containing $s$.  Because the integrations consist only of collapsing
vertices and edges and also making extra copies of edges, these $n$
legs must have come from $l$ legs in the original graph, where $l \le
n$.  In addition, because the process of integration does not change
the connectedness of the graph, when the $l$ legs in the original
graph are removed, it will break into two disjoint, connected graphs,
one containing $t$ and the other containing $s$.  An example of this
is given in Fig. 9.  We also note that since each graph is 1PI, to
break it into two we must remove at least 2 edges.

This all implies that the only graphs that can have a leading term of
$1/(t-s)^2$ are those that are broken into 2 when 2 legs are removed;
the only graphs that can have a contribution of $1/(t-s)^4$ are those
that are broken into 2 when 2, 3 or 4 legs are removed, and, in
general, only the graphs that can be broken into two connected pieces
when 2, 3, $\dots$, or $n$ legs are removed can contribute a term of
order $1/(t-s)^n$.  A simple counting argument shows that when $l$
legs are removed, the maximum net charge either of the 2 resulting
graphs can have is $l/2g$.  Because the net charge is always an
integer, when $l$ is equal to 2 (and g is greater than 1) this means
that the net charge must be zero.

Thus we can classify the graphs according to what their leading
behavior is, and we can determine which graphs will contribute to any
particular term in the expansion in Eq.(\ref{propsum}).  To make
contact with the previous subsection, we remark that for the insertion
of two charges, the only configuration that breaks into two graphs
when two lines are removed is precisely the one shown in Fig. 8b.

To arrive at a useful way of estimating graphs (which should also
apply in the non-equilibrium Minkowski space formalism), we observe
that every time we evaluate a residue of a pole at $t_k = s_j +
i\delta$, we are taking $t_k$ to be very close to $s_j$.  If we then
evaluate an $s_j = t_l +i\delta$ residue, we evaluate $s_j$ close to
$t_l$, so in turn that means $t_j$ is also close to $t_l$.  Following
through on this observation, we see that for the final result, all the
points are either evaluated close to $t$ or close to $s$, and whether
it is $t$ or $s$ depends on whether, when we remove the $n$ legs, the
point is in the graph connected to $t$ or to $s$.  Thus it appears
that the only contributions to the integral come from all the ways to
take some of the vertices close to $t$ and the remaining vertices
close to $s$.  The exponent of the leading contribution will then be
determined by the net charge of each of the two resulting subgraphs.
This is exactly what was done in Section V for the case of two
dipoles.

We now return to calculating the coefficient, $a_{N2}$, of the
$1/(t-s)^2$ part of the charge-charge correlator.  From the previous
discussion, we know that this should come from all ways of forming a
neutral multipole around $t$ and a neutral multipole around $s$.  As
long as $t-s >> \delta$, we can assume that all the charges in each
multipole are much closer to each other than $t$ and $s$ are to each
other.  We will let $t_0 \dots t_{m-1}$ and $s_0 \dots s_m$ be the
charges close to $t$ and $t_{m+1}\dots t_N$ and $s_{m+2}\dots s_N$ be
the charges close to $s$.  To simplify the notation, in most of what
follows we will let $t_m$ equal $t$ and $s_{m+1}$ equal $s$.  Next, we
will change variables so that $t_i = p_i + t$, $s_i = q_i + t$ for the
charges close to $t$ and $t_j = p_j + s$, $s_j = q_j + s$ for the
charges close to $s$.  Then the expression for $P$ becomes
\begin{equation}
P = P_1 P_2 I^g,
\end{equation}
where
\begin{equation}
P_1 = {\prod_{i,j=0;\, i<j}^m\left((p_i-p_j)^2 + \delta^2\right)^g
       \left((q_i-q_j)^2 + \delta^2\right)^g \over
       \prod_{i,j = 0}^m\left((p_i-q_j)^2 + \delta^2\right)^g}, \label{P_1}
\end{equation}
\begin{equation}
P_2 = {\prod_{i,j=m+1;\, i<j}^N\left((p_i-p_j)^2 + \delta^2\right)^g
       \left((q_i-q_j)^2 + \delta^2\right)^g \over
       \prod_{i,j = m+1}^N\left((p_i-q_j)^2 + \delta^2\right)^g},
\label{P_2}
\end{equation}
and
\begin{equation}
I = {\prod_{i=0}^m \prod_{j=m+1}^N\left((t-s+p_i-p_j)^2+ \delta^2\right)
     \left((t-s+q_i-q_j)^2+ \delta^2\right) \over
   \prod_{i=0}^m \prod_{j=m+1}^N\left((t-s+p_i-q_j)^2+ \delta^2\right)
     \left((t-s+q_i-p_j)^2+ \delta^2\right)}.  \label{Idef}
\end{equation}
$P_1$ and $P_2$ just look like the original integral, but for a
smaller number of charges, so they contain the intra-multipole
interactions.  The expression for $I$ contains all the interactions
between the two different multipoles.  In the numerator, the positive
and negative charges of the first multipole interact with charges of
the same sign in the second multipole, and in the denominator the
charges of the first multipole interact with the charges of opposite
sign in the second multipole.  Because the multipoles are both
neutral, and because every factor in Eqn. (\ref{Idef}) depends on
$t-s$, both the numerator and denominator have the same number of
factors of $t-s$.  Once we divide through by $t-s$, similar counting
tells us that the number of times $p_i/(t-s)$, $q_i/(t-s)$ and
$\delta^2/(t-s)^2$ each appear in the numerator equals the number of
times each of these appear in the denominator.  If we expand $I$ out
for large $t-s$ and count all the terms that contribute to order
$1/(t-s)^2$, we find
\begin{equation}
I = 1 + {1\over (t-s)^2}2\sum_{i,j}(p_i p_j + q_i q_j - p_i q_j - p_j q_i),
\end{equation}
where $p_i$ and $q_i$ run over all the charges in the first multipole and
$p_j$ and $q_j$ run over all the charges in the second multipole.
Then
\begin{equation}
I^g =  1 + {2g\over (t-s)^2}\sum_{i,j}(p_i p_j + q_i q_j - p_i q_j - p_j q_i).
\label{I^g}
\end{equation}
The important feature of the $1/(t-s)^2$ part is that it is odd under
changing the signs of the coordinates of all the charges in only one
multipole.  Meanwhile, $P_1 P_2$ is even under such a sign change, so
once we integrate over all the coordinates, the $1/(t-s)^2$ part
vanishes and we are left only with the $1/(t-s)^3$ (which should
vanish once we sum over all configurations of the charges) and the
$1/(t-s)^4$ parts.  Thus, the coefficients, $a_{2N}$ should vanish to
all orders in perturbation theory and the charge-charge correlation
functions, $\langle 0 | \Gamma e^{i\gamma\phi(t)}\Gamma^*
e^{-i\gamma\phi(s)}|0\rangle$, should go as $a_4\delta^2/(t-s)^4$, for
some constant $a_4$.  It is considerably more difficult to evaluate
this constant.

One final remark is that if we had regulated only the denominator,
then the previous argument would not have gone through: the
$\delta^2/(t-s)^2$'s from the denominator would no longer be canceled
by the $\delta^2/(t-s)^2$'s from the numerator, so that $a_{2N}$ would
be non-zero.  In this case, the correlation functions instead would go
as $1/(t-s)^2$.

\subsection{Implications for the Non-equilibrium Case}

Even for the non-equilibrium case, we can use our analysis of the
graphs in the preceding subsection to guide us in determining which
graphs should give the leading contributions to the current-current
correlation functions.  To calculate the singularity at $\omega=0$, we
can use the same neutral multipole expansion as in the end of the
previous section.  The only changes to Eqs. (\ref{P_1}, \ref{P_2}, and
\ref{Idef}) for the intra-multipole and inter-multipole interactions
are that we must now use the non-equilibrium regulators which depend
on the $\alpha_c(t_i, s_j)$'s.  Also, Eq. (\ref{P_1}) for the
multipole $P_1$ will now be multiplied by $\prod_{i=0}^m e^{i\omega_0
p_i} \prod_{j=0}^m e^{-i\omega_0 q_i}$ and Eq. (\ref{P_2}) for $P_2$
will be multiplied by $\prod_{i=m+1}^N e^{i\omega_0 p_i}
\prod_{j=m+1}^N e^{-i\omega_0 q_i}$.  Consequently, $P_1P_2$ no longer
remains unchanged when all the signs of the vertices in one multipole
are reversed.  Therefore, according to Eq. (\ref{I^g}) the
contribution to the current-current correlation function when one
multipole is close to vertex $t$ and the other is close to vertex $s$
goes as
\begin{equation}
1 + \int\prod_{k=0\atop k\ne m}^N dp_k \prod_{l=0 \atop l\ne m+1}^N dq_l
\, P_1 P_2 \sum_{i,j}(p_i p_j + q_i q_j - p_i q_j - p_j q_i)
{2g\over (t-s)^2},
\end{equation}
where $p_i$ and $q_i$ are in the first multipole and $p_j$ and $q_j$
are in the second multipole.  Also, we only take the connected graphs
in the multipoles $P_1$ and $P_2$.  Thus, to all orders in $\Gamma$,
the correlator goes as $1/(t-s)^2 + O\left({1/|t-s|^3}\right)$.  This
means that, at low frequency, the noise spectrum should have a
singularity that goes as $|\omega|$ at every order in $\Gamma$.  Here
we are assuming that for $g>1$ the neutral multipoles are all bound,
just as they are in the equilibrium case.

In the non-equilibrium case, we also expect singularities at $\omega =
\pm\omega_0$ and possibly also at $\omega = n \omega_0$ for other integer
values of $n$.  To find the leading behavior at these singularities we
use the fact that the expression for $P$ in Eq. (\ref{PKeldysh}) can
be expressed as a product $AB$.  As in the preceding subsection, $A$
is a determinant, and $B$ contains the corrections that naively go as
$\left(1 + O(\delta)\right)$. For the non-equilibrium case, $A$ was
defined at the beginning of this section as $\det M_{ij}$ where
\begin{equation}
M_{ij} = {1\over \delta \alpha_c(t_i, s_j) + i(t_i-s_j)}.
\end{equation}
The graphs for the $A$ defined here are identical to those in the
previous subsection, except for the choice of regulator. This means
that all of our previous counting arguments should apply.  However,
the form of $B$ is now much more complicated than before, so it is not
clear whether it modifies the counting in the same simple way as
before.  Because the expression for $P$ in Eq. (\ref{PKeldysh}) and
the expression for $A$ both contain the information about which branch
each charge is on, and since the only difference between the two
expressions is the choice of regulator, for convenience we will choose
to work with $A=\det M_{ij}$ instead of with $P$.  (In the equilibrium
case, we have seen that picking a different regulator does not change
the types of terms that can appear in the final answer; it just
changes the value of the coefficient in front of each term, possibly
setting some to zero.  In case of a discrepancy, the choice of
regulator should reflect the physics at hand, so it is useful to keep
in mind that in $P$ the interactions in the Coulomb gas are regulated
and in $A$ the fermion-like propagators in the matrix $M$ are
regulated.)

For $\det M_{ij}$, our counting and classification of graphs proceeds
as before.  This implies that if we can break the graph into two
connected multipoles with charge $Q$ and $-Q$, respectively, then the
graph will give a leading contribution of
\begin{equation}
a_Q {e^{iQ\omega_0 t} e^{-iQ\omega_0 s} \over (t-s)^{2Qg}}, \label{Qmult}
\end{equation}
as long as all charges within a multipole are close to one another.
This will give the singularity $|\omega \pm Q \omega_0|^{2Qg-1}$.  For
example, the graph in Fig. 9b will give a contribution as in Eq.
(\ref{Qmult}) with $Q=1$ and $g=2$.  Without performing the integral,
we cannot determine whether $a_Q$ (which can depend on $\delta$ and
$\omega_0$) is non-zero.  However, from this line of reasoning, we can
conclude that the $\Gamma \Gamma^* |\omega \pm
\omega_0|^{2g-1}$ singularity should only receive corrections that go at
least as $|\omega\pm \omega_0|^{2g-1}$ at all higher orders in $\Gamma
\Gamma^*$.  Similarly, at higher multiples of $\omega_0$ we expect the
singularities to be even smoother because they go at least as $|\omega
\pm Q\omega_0|^{2Qg-1}$.

As a check on these calculations, we note that we can apply the same
analysis of the graphs and similar counting arguments even at $g=1$.
In this case, every connected graph is just a simple polygon with
alternating charges at the vertices.  It is straightforward to see that
when any such graph is divided
into two disjoint, connected parts, each part can only have a total
charge of $0$ or $\pm1$, and exactly two lines must be cut.  Therefore,
the only singularities we can obtain are $|\omega|$ and $|\omega \pm
\omega_0|$, with no higher order corrections.  These results agree
with those in Ref. \cite{Yang} and give strong evidence that our method of
analyzing the graphs works even for the non-equilibrium case.

\section{Conclusion}

In this work we defined a framework for the study of equilibrium
and non-equilibrium noise in 1-D Luttinger liquids. The interactions
give rise to correlations that are manifest in the noise spectrum.
The correlations are responsible both for algebraic singularities in
the noise power spectrum and for the nonlinear dependence of the
strength of such singularities on either the applied voltage between
the terminals of the 1-D system or the temperature. The information
carried by both the form of the singularities and their strength can
help us identify Luttinger liquid states in experiments.

The picture of the tunneling in terms of the Coulomb gas (and its
dipole gas interpretation) is attractive because it gives us an
intuitive way to think about the tunneling in the Keldysh formalism.
This picture provides a unified description of the low and high
frequency noise: correlations between different dipoles define the
structure of the noise near zero frequency, whereas correlations
between the two charges within the dipole should contribute to the
noise near the ``Josephson'' frequency $\omega_J=e^*V/\hbar$.  Using
formal diagrammatic techniques we have justified this interpretation,
and, for integer $g$, we have obtained exact answers for the form of
the singularity in the equilibrium case.

One particulary striking result we obtained is that the form of the
leading singularity at zero frequency ($\propto |\omega|$) is the same
for strongly correlated Luttinger liquids as well as for
non-interacting systems. The effects of correlations in the case of
low-frequency noise is present only in the strength of the
singularity, with a strong non-linear dependence on the applied
voltage that is proportional to $V^{4(g-1)}$.

Although our Coulomb gas picture and the accompanying formalism has
enabled us to calculate the form of the singularities to all orders in
perturbation theory, beyond the order $|\Gamma|^4$ it is too
cumbersome to find the strength (i.e. the coefficient in front) of
these singularities.  We would also like to point out that the
structure of the noise far away from the frequencies $n\omega_J$, at
higher orders in perturbation theory, is unknown; the information we
are able to obtain is limited solely to frequencies near the singular
points. One exception is the exactly solvable case of $g=1$, where we
find that the noise spectrum must have the form $a +b |\omega| + c
|\omega \pm \omega_J|$, where $a$, $b$ and $c$ can be calculated from
the non-equilibrium voltage and the transmission coefficient. Thus in
this case we recover the results for non-interacting electrons.
Indeed, the framework we presented can be used with $g=1$ for studying
coherence effects which appear in the noise for non-interacting
electrons and are due to the Pauli principle, because the statistics
enter in the formulation we use through the bosonization.

There are two points in this work that need further exploration. The
first is the apparent fine point of better understanding the role of
the short distance cutoff in our calculations.  We need either to
determine whether the non-equilibrium noise is sensitive to our choice
of regulator or else to show that our choice of regulating the
fermion-like propagators instead of the Coulomb gas is the physical
one.  The second, and more important, question is to understand
non-pertubative effects. For example, one expects that the position
of finite voltage singularities should depend on $\Gamma$. In the case
of tunneling between edge states, when we increase the current, the
frequency should shift from $e\frac{V}{\hbar}$ to
$\frac{e}{m}\frac{V}{\hbar}$ as we go from the configuration in Fig.
1c, where the electrons are tunneling, to the one in Fig. 1b, where
the quasi-particles are tunneling.  This is not reflected in our
perturbative calculations.  However, we have evidence that within our
Coulomb gas picture this shift can be explained by non-perturbative
effects, and we hope to address this issue in a future paper.

\

\begin{center}
{\bf ACKNOWLEDGEMENTS}
\end{center}

This work is supported by the NSF Grant No. DMR-91-14553. D. F. would like
to thank the M.I.T. Center for Theoretical Physics for their hospitality.

\newpage

\begin{center}
{\large FIGURE CAPTIONS}
\end{center}

\

Figure 1. Schematic drawing of the geometries for tunneling in 1-D
Luttinger liquids.  A channel connected to two reservoirs is shown in
(a), with a potential barrier or weak link in the middle.  The
geometries for tunneling between edge states are shown in (b) and (c).
By adjusting the gate voltage $V_G$ one can obtain either a simply
connected QH droplet (b), or two disconnected QH droplets (c).  For
the geometry in (b) both electrons and quasiparticles (carrying
fractional charge) can tunnel from one edge to the other, whereas for
the geometry in (c) only electrons can tunnel. The tunneling current
$I_t$ depends on the applied voltage between the right and left edges.

\

Figure 2. An insertion of an operator $e^{+i\gamma \phi(t)}$
correspond to the insertion of a charge $+$ on the contour at time
$t$.  Similarly, an insertion of an operator $e^{-i\gamma \phi(t)}$
correspond to an insertion of a charge $-$ at time $t$. The time $t$
is ordered along the contour shown, and there is a distinction between
charges placed on the top and bottom branches. For illustration, in
the example shown we have for the number of $+$ and $-$ charges in the
$t$ and $b$ branches $Q_+^t=3$, $Q_-^t=2$, $Q_+^b=2$ and $Q_-^b=3$.
Only terms that have zero total charge $Q=Q_+^t+Q_+^b-Q_-^t-Q_-^b$ can
give a non-zero contribution to $Z$.

\

Figure 3. The applied voltage $V$ between the terminals or edges
creates an unbalance of charge between the top and bottom branches.
Since $+$ and $-$ charges correspond respectively to tunneling from
$R\rightarrow L$ and $L\rightarrow R$, an excess of charge in the top
branch correspond to net tunneling in one direction.

\

Figure 4. The charges that form the Coulomb gas can form a dipole
phase.  In this phase, the expression for $Z$ can be recast as a sum
over dipole strengths and positions, instead of summing over the
locations of the $+$ and $-$ charges.

\

Figure 5. The four types of dipole, classified according to the
position of the $+$ and $-$ charges comprising it. In the $c_+$ dipole
the $+$ charge is on the top branch and the $-$ charge is on the
bottom. In the $c_-$ the $-$ charge is on the top and the $+$ is on
the bottom. In the $t$ dipole both charges are on the top branch, and
in the $b$ dipole both charges are on the bottom branch. Notice that
only the $c_\pm$ dipoles contribute to a net current, as they create
an unbalance of charge between the top and bottom branches. The $t$
and $b$ dipoles contribute to the noise, but not to the current.

\

Figure 6. Two dipoles will interact because of the relative position
between the charges that comprise them. The figure shows two dipoles
with center of mass positions $t_1$ and $t_2$ and strengths $p_1$ and
$p_2$.

\

Figure 7. The expression for the correlation between many charges can
be expressed as a power of the determinant of a matrix $M$. The matrix
element $M_{ij}$ can be represented diagrammatically as a line
connecting a $+$ charge at position $t_i$ to a $-$ charge at position
$s_j$, as shown in (a). The determinant contains different ways of
pairing the charges (b). Finally, when raising the determinant to the
power $2g$ (done in this figure for $g=1$), we generate different ways
of connecting the charges such that exactly $2g$ lines leave each $+$
charge and exactly $2g$ lines arrive at each $-$ charge, as shown in
(c).

\

Figure 8. The graph corresponding to independent dipoles is shown in (a),
with all lines leaving the $+$ charge arriving at the $-$ charge (here
we use $g=2$ for illustration). One of the graphs contributing to the
dipole-dipole correlation is shown in (b). Each leg connecting the two
dipoles contributes to order $1/t$, so that the dipole-dipole correlation
is of order $1/t^2$.

\

Figure 9.  A sample graph with $g=2$ that gives a contribution of
$1/(t-s)^4$ after it is integrated.
The final graph with 4 legs is shown in (a).  It is obtained by integrating
over the vertices $t_1$, $t_2$, $s_1$ and
$s_2$ in the graph shown in (b).  The 4 final legs in the final graph
come from the 4 bold-faced legs.  In (c), the two disjoint graphs
(or multipoles) resulting from removing the 4 bold-faced legs are shown.

\end{document}